\documentclass[10pt,a4paper]{iopart}

\usepackage{iopams}  
\usepackage[breaklinks=true,colorlinks=true,linkcolor=blue,urlcolor=blue,citecolor=blue]{hyperref}

\usepackage[pdftex]{graphicx}  

\newcommand{\pos}[1]{\mathbf{r}_{#1}}
 
\begin{document}
 
\title{Electron localisation in static and time-dependent one-dimensional model systems}
\date{\today}
\author{T.\ R.\ Durrant$^1$\footnote{Present Address: Department of Physics and Astronomy and London
 Centre for Nanotechnology, University College London, Gower Street, London, WC1E 6BT,
 United Kingdom}, M.\ J.\ P.\ Hodgson$^1$\footnote{Present Address: Max-Planck-Institut f\"ur Mikrostrukturphysik,
 Weinberg 2, D-06120 Halle, Germany}, J.\ D.\ Ramsden$^1$, R.\ W.\ Godby$^1$}
\address{$^1$ Department of Physics, University of York, and European Theoretical Spectroscopy Facility, Heslington, York YO10 5DD, United Kingdom} 
 
\begin{abstract}
The most direct signature of electron localisation is the tendency of an
 electron in a many-body system to exclude other same-spin electrons from
 its vicinity. By applying this concept directly to the exact many-body wavefunction, we find that localisation can vary considerably between different ground-state systems,
 and can also be strongly disrupted, as a function of time, when a system is driven by an applied electric field. We use this measure to assess the
 well-known electron localisation function (ELF), both in its approximate single-particle form 
 (often applied within density-functional theory) and its full many-particle form. The full ELF always gives an excellent description of localisation, 
 but the approximate ELF fails in time-dependent situations, even when the exact Kohn-Sham orbitals are employed.
\end{abstract}
 
\submitto{\JPCM}
\maketitle \ioptwocol

\section{Introduction}

Density functional theory (DFT) \cite{hohenberg1964inhomogeneous} replaces the many-body (MB) wavefunction with
 the electron density as its fundamental variable, making electronic structure problems computationally tractable.  
 However, this transformation provides new challenges, as in the non-interacting world of Kohn-Sham (KS) electrons, the many-body
 behaviour of electrons is concealed within the exchange-correlation (xc) potential \cite{PhysRev.140.A1133}.
 
Electron localisation, describing the tendency of an electron to exclude other same-spin electrons from its vicinity (i.e. position entanglement of like-spin electrons), is one such property \cite{becke1990simple, savin1997elf}. Although a commonly used concept, it is not always well defined. We choose to start from the idea that localised electrons tend to avoid one another, whereas delocalised electrons will share the same region of space.
Localisation is partly driven by Pauli exclusion, which acts to localise like-spin electrons in separate regions. The Coulomb interaction further enhances the tendency to localise \cite{dobson1991interpretation}.
The electrons' attempts to avoid each other increase the kinetic energy, which, however, is minimized by spreading electrons over as large a volume as possible. It is the balance between these factors that makes electron localisation challenging to quantify. 
 
An understanding of electron localisation is useful chemically, placing the ubiquitous concepts of chemical bonds and localised electron pairs on a formal footing \cite{silvi1994classification}.  Although measures of localisation provide an understanding of bonds and electron pairs made up of electrons of \emph{opposite spin}, these details are revealed by looking at the localisation of \emph{like-spin} electrons, which provides the regions in which localised opposite-spin electron pairs can be found. Localisation also describes a fundamental aspect of electron correlation that  approximate DFT functionals should take into account  \cite{LOCFUNC,KOHNAIRY,hao2014using,PhysRevLett.115.036402,PhysRevLett.100.146401,PhysRevLett.100.146401}.

We stress that localisation is a true \textit{many-body} property of electrons, dependent on the positions of all electrons, and not accessible through the spatial character of any one KS orbital. Although many early efforts focused on the extent of molecular or KS orbitals, it was soon realized that these orbitals are not unique and that quite different choices could be selected \cite{savin1997elf}.
 
In this paper, we explore the variation in localisation for a range of ground-state and time-dependent systems. 
To assess the merits of different descriptions of localisation, we introduce a comprehensive measure, calculated from the exact many-body wavefunction. 
Using this, we evaluate the reliability of the usual formulation of the electron localisation function (ELF), often used to characterize electron localisation in DFT calculations \cite{savin1997elf}.
 We find that this ELF correctly describes localisation across a range of ground-state systems, but fails in time-dependent
 situations where strong delocalisation occurs due to electron collisions. These errors are not present, however, if the ELF is formulated from the true pair density.
 
\section{Measures of localisation}

To calculate our measure, we note that in localised systems each electron will avoid the others by tending to stay in its own region of space. We first partition the system into distinct regions, each containing charge corresponding to one electron, and examine the actual distribution of the $N$ electrons among these regions. To simplify our definition, we consider a wavefunction with only one type of spin. The probability that each of the $N$ electrons is in a \textit{different} region is calculated using a mask function $M$. $M(r_1,\dots,r_N)$ is defined as 1 where $r_1,\dots,r_N$ all lie in distinct regions, and 0 otherwise, and interrogates the probability of such a situation via the $N$-electron wavefunction $\Psi$. The probability is then given by
\begin{equation}
p =  \int M \! \left( \pos{1}, \dots, \pos{N} \right) \, \left| \Psi \! \left( \pos{1},\dots,\pos{N} \right) \right|^2 d^3\pos{1}  \dots  d^3\pos{N} \, .
 \label{relmdef}
\end{equation}
The inclusion of spin replaces the wavefunction with the appropriate \emph{N}-body same-spin density matrix. 
 
Equation~\ref{relmdef} fully defines our measure once a set of one-electron
regions is provided via $M$. 
In the
present paper we study one-dimensional finite systems,
where the one-electron regions in $M$ are straightforwardly identified from
the cumulative integral of the electron density: starting from one edge, the system is simply divided into $N$ regions, each containing the charge of one electron.
More generally, the set of regions which most faithfully
represents how the electrons are localised in the system would be that which
maximises Eq.~\ref{relmdef} \footnote{In three dimensions, this remains a complex task, but,
pragmatically, the choice of regions can be advised by physical
considerations: the locations of minima in the charge density, for
instance, provide suggestions for the edges of these regions \cite{PhysRevB.93.155146}.}.
 
To form our regional electron localisation measure (RELM), we scale $p$ with reference to the probability $p_0=N! / N^N$ of finding exactly one electron in each region for an ideal delocalised and uncorrelated state:
\begin{equation}
\mathrm{RELM} = \frac{p -p_{0}}{1 - p_{0}}.
\label{}
\end{equation}
This makes our measure $0$ when a system is ideally delocalised, and 1 when fully localised. As the probability of simultaneously finding only one electron in each region increases, so does our measure of localisation. As with the ELF, our measure is calculated for each spin index independently.
 
Of course, in most cases the MB wavefunction, on which the RELM relies, is too expensive to calculate. The traditional method of approximating the localisation in a system is the ELF of Becke and Edgecombe \cite{becke1990simple}. The following reference provides a comprehensive review of the ELF \cite{savin1997elf}. Originally developed for Hartree-Fock (HF) calculations, the method can also be applied
 to Kohn-Sham orbitals. The ELF is based on the quantity $D_\sigma$, defined by the Taylor series
 \begin{equation}
\rho^{\sigma  \sigma}_{\mathrm{Cond}} \left( \mathbf{r}, s\right) =  \frac {{D_\sigma  \left( \mathbf{r} \right)}  s^2}{d} + O(s^3) ,\label{taylor}
\end{equation}
where  $\rho^{\sigma \sigma}_{\mathrm{Cond}}$ is a conditional probability, the  probability given that an electron has been found at position $\pos{}$  that a second electron will be found at distance $s$ from this position.  $d$
is the spatial dimensionality of the system and $\sigma$ is a spin index. $D_\sigma$ characterises the probability of finding a second (same-spin) electron very close to the reference electron and is a measure of localisation in its own right.
 
Later work by Dobson \cite{dobson1991interpretation} provides an equivalent  definition of  $D_\sigma \left( \mathbf{r} \right)$ to that given in  Eq.~\ref{taylor},  allowing this quantity to be calculated directly from  the same-spin pair density,  using the equation
 \begin{equation}
D_\sigma  \left( \mathbf{r} \right) = \frac { \left[  \nabla^{2}_{\pos{}'}  n_{2}^{\sigma}(\mathbf{r} , \pos{}')   \right]_{\pos{}'= \mathbf{r}}} {2  n_{\sigma}(\mathbf{r})},
 \label{ELFExact}
\end{equation}
where $n^{\sigma}_{2}(\mathbf{r} , \pos{}')$ is the pair density. Using this expression, it is possible to calculate the ELF directly from the wavefunction; we term this the ``exact ELF''.
 
As $D_\sigma$ has a strong dependence on the local density, it is not easily interpreted directly. To produce the ELF, Becke and Edgecombe scaled $D_\sigma$  as
 \begin{equation}
\mathrm{ELF}  \left( \mathbf{r} \right) = \frac{1}{1 + \left( \frac{D_\sigma \left(  \mathbf{r} \right)} {D_{\sigma, \mathrm{H}} \left( \mathbf{r} \right)  }\right)^2 }.
\label{ELF}
\end{equation}
This expression compares the local value of $D_\sigma$ with that of the Hartree-Fock homogeneous electron gas of the same local density, $D_{\sigma, \mathrm{H}}$ (a convenient
reference system against which to compare the actual value of $D_\sigma$). Hence, ELF ranges from zero to one, where 1 represents total localisation and 0.5 represents the degree
of localisation in a HEG of the same density.

In most systems, the pair density is not available, so the exact definition  of Eq.~\ref{ELFExact} can not be used. Becke and Edgecombe found an approximate expression to calculate $D_\sigma$ in terms of single-particle orbitals $\phi$:
\begin{equation}
{D_\sigma} {\approx} \sum_i^{N_\sigma} {|\nabla \phi^{\sigma}_{i}|}^2 - \frac{1}{4} \frac {{|\nabla n_\sigma |}^2} {n_\sigma},
\label{ELFApprox}
\end{equation}
where $n_\sigma$ is the electronic density. We use Hartree atomic units (a.u.) here and henceforth. This equation is exact in a HF treatment, and is also commonly applied to KS orbitals \cite{savin1992electron}. We term
this the ``approximate ELF''.
 
As shown, Becke and Edgecombe's approach relies on two main assumptions: that $D_\sigma \left( \mathbf{r} \right)$ is an effective local description of localisation, and that the approximation to it given by Eq.~\ref{ELFApprox} is satisfactory.
If either of these assumptions fail, the ELF will give a misleading picture of electronic behaviour. ELF calculations
have been widely used, but less is known about their accuracy. Previous work on molecules has suggested that ELF can perform poorly for DFT calculations where correlation is strong \cite{ELFCI}, by comparison with the accurate configuration interaction (CI) method. It remains unclear if this is caused by approximate ELF's reliance on a single Slater determinant, or is instead due to approximate exchange-correlation potentials leading to an incorrect degree of localisation.
 
ELF is a local measure of localisation and in order to look at the localisation of systems as a whole we define the average ELF, weighted according to the density,
$
\left< \mathrm{ELF} \right> = \frac {1}{N} \int \mathrm{ELF} \left( \pos{} \right) n \left( \pos{} \right) d^3\pos{} \, ,
$
which can take values between zero and one. 
 
\section{Calculations}

\subsection{Approach}

All the results presented in this paper were calculated using the iDEA code \cite{hodgson2013exact}. The Schr\"odinger equation is solved exactly for 1D systems of spinless electrons to find the exchange-antisymmetric MB wavefunction, both for ground-state
and for time propagation after an electric field is applied. This provides the charge density for all our model systems. As is appropriate in 1D, we use a softened Coulomb interaction \cite{hodgson2013exact}.
Spinless electrons maximise the richness of exchange and correlation for a given computational effort, as each KS electron occupies a distinct KS orbital. From the MB density, the exact KS orbitals that reproduce the density
are calculated \cite{hodgson2013exact}.

When using DFT to evaluate properties, it is often unclear which errors are intrinsic to the method being applied and which are caused by failures of the xc potential in the underlying DFT calculation. The failure of common xc potentials to adequately localise electrons is well known \cite{kummel2008orbital,PhysRevLett.100.146401,cohen2008insights}. Access to the exact KS orbitals enables us to assess these approximations directly and understand which methodologies would fail even if the universal functional was known.

When forming the ELF it is important that the reference system be chosen to have the correct characteristics, as noted in \cite{rasanen2008electron}. Specifically, the $D_{\sigma, \mathrm{H}}$
 function for the Hartree-Fock homogeneous gas of spinless electrons in one dimension is $\frac{1}{6} \pi^2 n^3$. In an earlier paper a similar ELF for one-dimensional systems, constructed in an analogous 
manner, was applied to coupled electron-nuclear wavefunctions \cite{erdmann2004time}. The ``localisation regions'' used to define our RELM localisation measure are (as noted above) straightforward, though it is, of course, essential to update them as a function of time in a time-dependent system.
 
 
\subsection{Ground-state localisation}

\begin{figure}[htbp]
  \centering
  \includegraphics{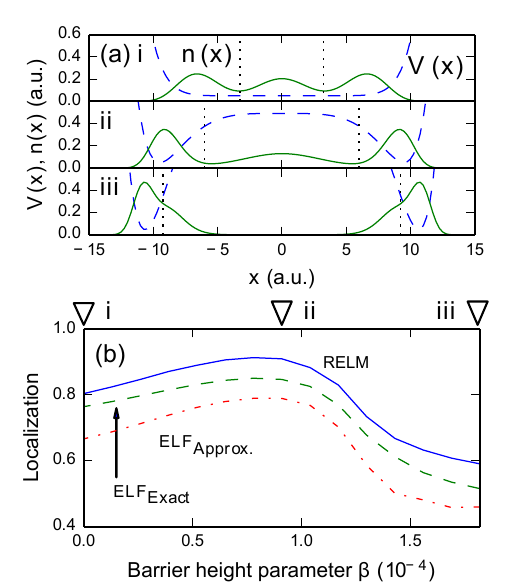}
  \caption{\textit{3-electron double wells} --- (a) Plots of the external potentials (dashed blue) of three selected wells as
  the barrier height is increased. The ground-state charge densities (solid green) of these potentials are shown. The localisation
  regions used in the RELM calculations are also shown (dotted black); each contains exactly one electron's worth of charge. (b) The localisation
  of the family of potentials, calculated using the three methods introduced -- RELM, exact average ELF and approximate average ELF.
  All three measures agree how the barrier influences localisation. Triangles indicate the values of $\beta$ for which the potentials are plotted in (a).}
\label{DoubleWell}
\end{figure}

First we look at a family of three-electron double-well potentials and calculate the localisation of their ground states. These external potentials are defined as
\begin{equation}
V\left(x\right) = \alpha x^{10} - \beta x^4 .
\label{DoubleWellPot}
\end{equation}
A constant value of $\alpha=5 \times 10^{-11}$ a.u. was used, while the value of $\beta$ was varied:
 when $\beta=0$ there is no barrier in the potential, and as it is increased the height of the barrier grows. Three examples of the potentials and densities for different
 values of $\beta$ are shown in Fig.~\ref{DoubleWell}. This family of wells is interesting as the double-well potential only provides two natural sites
 for the three electrons to occupy.
 
Fig.~\ref{DoubleWell}(a) shows some of the ground-state potentials that make up this family and the effects of the potential barrier on the electrons. At first, the middle electron stays in the barrier region owing to the Coulomb repulsion, as its height increases in (i) to (ii). As the presence of the barrier disperses the central electron across the system, it also drives the outer electrons toward the boundaries of the system, acting to {\it increase} the localisation of the system rather than decrease it (see Fig.~\ref{DoubleWell}(b)). As the strength of the barrier is increased further, it becomes energetically favourable for the central electron to move into the two side wells, as in (iii), reducing localisation. 
 
In Fig.~\ref{DoubleWell}(b), the three measures agree on how the localisation of this family of potentials varies. The similarity between RELM and the exact average ELF is striking as the two methods are based on different mathematical
interpretations of localisation: RELM is scaled probabilistically and ELF is scaled with respect to the HEG as reference. Approximating $D_\sigma$ does lead to a systematic lowering of the calculated $\left<\mathrm{ELF}\right>$, but still yields the correct trend across the range of localisations calculated. 
 
\begin{figure}[htbp]
  \centering
  \includegraphics{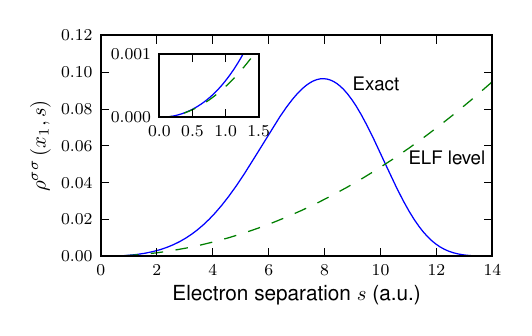}
  \caption
  {A plot of the conditional probability $\rho^{\sigma \sigma}_{\mathrm{Cond}} \left( x_1, s\right)$ against electron separation $s$ where $x_1$ is fixed at the density maximum (2.64 a.u.) for the two-electron ground state of the potential of Fig.~\ref{DoubleWell}(a)(i). The conditional probability (blue solid) is plotted with the ELF level approximation to it (green dashed). Insert: magnification of short $s$ behaviour. $D_\sigma$ does not contain any long range information and only correctly characterises short distances, in this example only $\sim0.6$ a.u. Our strong ELF results suggest that this neglected long-range behaviour is not important for localisation.
  }
\label{TaylorPlot}
\end{figure}

We next investigate the information contained in $D_\sigma$ in more detail. Fig.~\ref{TaylorPlot} shows an example plot for a two electron system (specified in
the caption), demonstrating the approximation made in Eq.~\ref{taylor} in practice. The conditional probability $\rho^{\sigma\sigma}_{\mathrm{Cond}}$ is plotted for a chosen value of $\pos{}$. As demanded by the Pauli exclusion principle, the probability of finding a second electron at the same position ($s=0$) is zero. The function then shows a peak at the most likely separation the second electron is found at (we would normally expect as many peaks as there are remaining electrons in the system).
 
Also shown on the plot is the approximated version of this conditional probability that is used in ELF calculations. As shown in Eq.~\ref{taylor}, the ELF approximates this conditional probability as $D_\sigma \left( \pos{} \right) s^2 / d$ and this is shown in the plot where $D_\sigma$ has been calculated using Eq.~\ref{ELFExact}. As shown in the inset in Fig.~\ref{TaylorPlot}, this approximation is only effective over very short electron separations. RELM calculations use information over all $s$, so the agreement between our ELF and RELM calculations suggests that this longer-range $s$ behaviour is not an important ingredient for a localisation measure to contain.
 
\subsection{Time-dependent localisation}

Next we look at a time-dependent system. As first derived by Dobson \cite{dobson1993alternative}, in the time-dependent regime the approximate ELF is modified by the addition of an extra term to Eq.~\ref{ELFApprox}, producing the time-dependent ELF (TDELF) \cite{burnus2005time}. This equation becomes
\begin{equation}
{D_\sigma} {\approx} \sum_i^{N_\sigma} {|\nabla \phi^\sigma_{i}|}^2 - \frac{1}{4} \frac {{|\nabla n_\sigma |}^2} {n_\sigma} - \frac{{j_\sigma}^2}{n_\sigma},
\label{TDELF}
\end{equation}
where $j_\sigma$ is the current density. 
 
For time-dependence, we return to the potential well having $\beta=0$, shown in Fig.~\ref{DoubleWell}(a)(i), and this time place two electrons in it. As before, we find the ground-state wavefunction of the well. Then at time $t=0$ we apply a strong uniform d.c. electric field (potential $-0.1x$), driving the electrons strongly towards the right-hand well, causing them to ``collide''. We also look at the non-interacting system with Pauli exclusion but no Coulomb interaction. We note that Coulomb interaction enhances localisation, and if the interaction strength is artificially enhanced the system is driven towards total localisation (RELM=1).
\begin{figure}[htbp]
  \centering
  \includegraphics{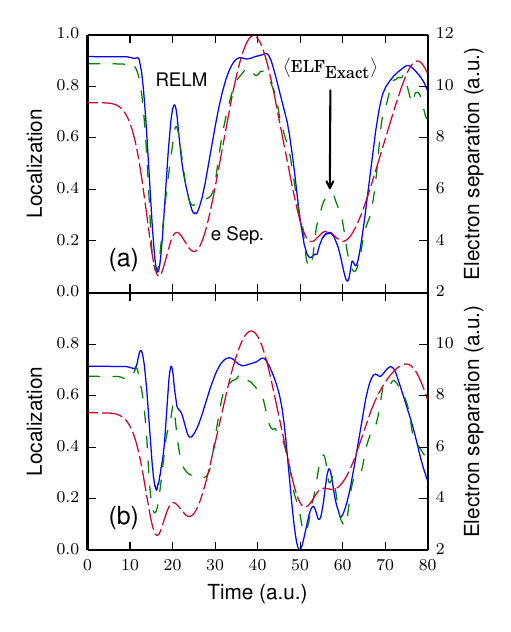}
  \caption{Localisation measures as a function of time for two electrons in the perturbed $x^{10}$ well of Fig.~\ref{DoubleWell}(a)(i). (a) The interacting system. RELM (solid blue) and exact average ELF (dashed green) continue to show good agreement. The expectation value of the electron separation (long dashed red) is also shown on the second axis and shows similar features. (b) The same plot for the non-interacting system. The behaviour of the system is similar, showing that Pauli exclusion is the main driver of localisation. RELM and ELF differ more significantly}
\label{TimeLoc}
\end{figure}
 
Fig.~\ref{TimeLoc}~(a) shows how the localisation of the interacting electrons changes during the 80 a.u. of the simulation. Broadly, these results show strong changes in localisation over time. This is in contrast to the notion that localisation is a persistent characteristic of a system. 
 
On both plots the expectation value of the electron separation $\left<  \hat{s}\right>$ is shown. The large localisation drops occur  around the two electrons becoming closer together. (These rapid drops in  localisation
also  appear in other systems that we have investigated and seem to be a common feature of electron collisions.) For our systems, $\left<  \hat{s}\right>$ seems to be a fair indicator of localisation in its own right.
 
Fig.~\ref{TimeLoc}(b) shows the calculation repeated with no Coulomb interaction. The behaviour is broadly similar,  but interaction seems to exaggerate some features and suppresses others.  This comparison strengthens the argument that Pauli exclusion is the  main driver of localisation. 
 
Again, for both plots RELM and exact ELF show the same trends, although both measures show some unique features. The close agreement
between the two measures still holds in a time-dependent context. We also note the agreement between exact ELF and RELM is reduced when the Coulomb interaction is turned off.
 
\begin{figure}[!]
  \centering
  \includegraphics{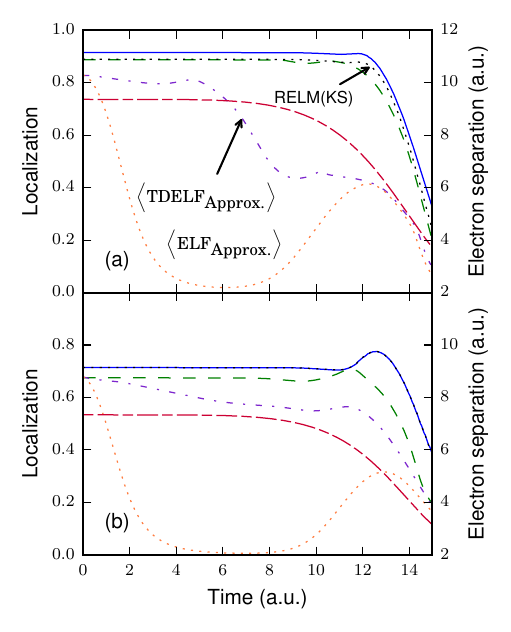}
  \caption{The curves in Fig.~\ref{TimeLoc} for a shorter time interval, with the addition of approximate-ELF calculations based on the exact KS potential. For the interacting system in (a), the average approximate TDELF (dotted-dashed purple) and the GS approximate ELF (dotted orange) both delocalise the electrons too soon. If RELM is calculated from a KS Slater determinant (dotted black) its value is slightly underestimated. The TDELF, though an improvement on the original approximation, has a large spurious drop in localisation around $t=5$ a.u. As shown in (b), this approximation performs better without interaction as the problematic drop is weaker. Without interaction, calculating RELM from a KS Slater determinant is exact and the two RELM curves coincide.}
\label{TimeLocSmall}
\end{figure}
 
Fig.~\ref{TimeLocSmall} shows approximate ELF calculations for the same systems: both the original formulation and the time-dependent extension. (These are shown for a shorter time interval due to the numerical challenge in calculating exact KS orbitals for 
later timesteps.) For the interacting system, the original (time-independent) ELF formulation performs very poorly and for most of the calculation shows the system as erroneously
delocalised. The extra current term in the TDELF makes a significant improvement. At around 5 a.u., however, it too shows an unphysical drop in localisation which analysis shows to occur in the left-hand well, predominantly associated with a large increase of the first term of Eq.~\ref{TDELF} in the left half of the system which is not compensated for by the other terms. Instead, the real
system delocalises both electrons at 12 a.u., later in the simulation.
Clearly, the TDELF's ignorance of correlation (beyond Pauli exclusion) in the wavefunction is limiting its description of localisation.
 
The approximate TDELF performs better when there is no electron-electron interaction [Fig.~\ref{TimeLocSmall}(b)]. It still shows a slow drop when the localisation is staying constant, but this erroneous drop is significantly weaker. 

Additionally, we calculate RELM from a Slater determinant of the KS orbitals, the wavefunction of the fictitious non-interacting KS electrons. The strong correspondence again demonstrates that Pauli exclusion is the main driver of localisation. This approximation is exact when there is no electron interaction, but is slightly weaker when the electrons are interacting. This approximation is more successful than the approximate ELF calculations, which assume that the KS orbitals obey the HF equations.
 
One failure of the present approximation, concealed in the definition of ELF, is that it is not positive definite, leading to non-physical negative values of $D_\sigma$. If these values are set to zero, a small improvement in accuracy is achieved. Negative values should serve as a warning that the method is not performing reliably \footnote{Very-low-density regions are generally unreliable for the approximate ELF and negative values are routinely seen here.}.
 
\section{Conclusions}

We have studied electron localisation, which provides insight into important aspects of many-electron correlation, using a variety of measures across a range of ground-state and time-dependent systems. Our results show the strength of the ELF approach, despite its focus on short range exclusion. We further find that the usual approximate ELF provides good results for a range of ground-state systems, notwithstanding its simplicity and neglect of correlation, allowing the extraction of physical meaning from a simple measure based on one-electron wavefunctions. 
In contrast, time-dependent systems can often become surprisingly delocalised as electrons collide with one another, and in this case the simple approximate ELF is no longer adequate. When many-electron excited states are being strongly explored, improved approximate localisation measures are required.

 \section{Acknowledgements}

We acknowledge funding from EPSRC, and thank David Tozer for helpful discussions.

\bibliographystyle{unsrt}
\bibliography{ref}

\begin{thebibliography}{10}

\bibitem{hohenberg1964inhomogeneous}
P~Hohenberg and W~Kohn.
\newblock Inhomogeneous electron gas.
\newblock {\em Phys. Rev.}, 136(3B):B864, 1964.

\bibitem{PhysRev.140.A1133}
W~Kohn and LJ~Sham.
\newblock Self-consistent equations including exchange and correlation effects.
\newblock {\em Phys. Rev.}, 140:A1133--A1138, Nov 1965.

\bibitem{becke1990simple}
AD~Becke and KE~Edgecombe.
\newblock A simple measure of electron localization in atomic and molecular
  systems.
\newblock {\em J. Chem. Phys.}, 92(9):5397--5403, 1990.

\bibitem{savin1997elf}
A~Savin, R~Nesper, S~Wengert, and TF~F{\"a}ssler.
\newblock {ELF}: The electron localization function.
\newblock {\em Angew. Chem., Int. Ed.}, 36(17):1808--1832, 1997.

\bibitem{dobson1991interpretation}
JF~Dobson.
\newblock Interpretation of the {Fermi} hole curvature.
\newblock {\em J. Chem. Phys.}, 94(6):4328--4333, 1991.

\bibitem{silvi1994classification}
B~Silvi and A~Savin.
\newblock Classification of chemical bonds based on topological analysis of
  electron localization functions.
\newblock {\em Nature}, 371(6499):683--686, 1994.

\bibitem{LOCFUNC}
MJP Hodgson, JD~Ramsden, TR~Durrant, and RW~Godby.
\newblock Role of electron localization in density functionals.
\newblock {\em Phys. Rev. B}, 90(24):241107, 2014.

\bibitem{KOHNAIRY}
W~Kohn and AE~Mattsson.
\newblock Edge electron gas.
\newblock {\em Phys. Rev. Lett.}, 81(16):3487, 1998.

\bibitem{hao2014using}
F~Hao, R~Armiento, and AE~Mattsson.
\newblock Using the electron localization function to correct for confinement
  physics in semi-local density functional theory.
\newblock {\em J. Chem. Phys.}, 140(18):18A536, 2014.

\bibitem{PhysRevLett.115.036402}
J~Sun, A~Ruzsinszky, and JP~Perdew.
\newblock Strongly constrained and appropriately normed semilocal density
  functional.
\newblock {\em Phys. Rev. Lett.}, 115:036402, Jul 2015.

\bibitem{PhysRevLett.100.146401}
P~Mori-S\'anchez, AJ~Cohen, and W~Yang.
\newblock Localization and delocalization errors in density functional theory
  and implications for band-gap prediction.
\newblock {\em Phys. Rev. Lett.}, 100:146401, Apr 2008.

\bibitem{PhysRevB.93.155146}
MJP Hodgson, JD~Ramsden, and RW~Godby.
\newblock Origin of static and dynamic steps in exact {Kohn}-{Sham} potentials.
\newblock {\em Phys. Rev. B}, 93:155146, Apr 2016.

\bibitem{savin1992electron}
A~Savin, O~Jepsen, J~Flad, OK~Andersen, H~Preuss, and HG~von Schnering.
\newblock Electron localization in solid-state structures of the elements: the
  diamond structure.
\newblock {\em Angew. Chem., Int. Ed.}, 31(2):187--188, 1992.

\bibitem{ELFCI}
E~Matito, B~Silvi, M~Duran, and M~Sol{\`a}.
\newblock Electron localization function at the correlated level.
\newblock {\em J. Chem. Phys.}, 125(2):024301, 2006.

\bibitem{hodgson2013exact}
MJP Hodgson, JD~Ramsden, JBJ Chapman, P~Lillystone, and RW~Godby.
\newblock Exact time-dependent density-functional potentials for strongly
  correlated tunneling electrons.
\newblock {\em Phys. Rev. B}, 88(24):241102, 2013.

\bibitem{kummel2008orbital}
S~K{\"u}mmel and L~Kronik.
\newblock Orbital-dependent density functionals: Theory and applications.
\newblock {\em Rev. Mod. Phys.}, 80(1):3, 2008.

\bibitem{cohen2008insights}
AJ~Cohen, P~Mori-S{\'a}nchez, and W~Yang.
\newblock Insights into current limitations of density functional theory.
\newblock {\em Science}, 321(5890):792--794, 2008.

\bibitem{rasanen2008electron}
E~R{\"a}s{\"a}nen, A~Castro, and EKU Gross.
\newblock Electron localization function for two-dimensional systems.
\newblock {\em Phys. Rev. B}, 77(11):115108, 2008.

\bibitem{erdmann2004time}
M~Erdmann, EKU Gross, and V~Engel.
\newblock Time-dependent electron localization functions for coupled
  nuclear-electronic motion.
\newblock {\em J. Chem. Phys.}, 121(19):9666--9670, 2004.

\bibitem{dobson1993alternative}
JF~Dobson.
\newblock Alternative expressions for the {Fermi} hole curvature.
\newblock {\em J. Chem. Phys.}, 98(11):8870--8872, 1993.

\bibitem{burnus2005time}
T~Burnus, MAL Marques, and EKU Gross.
\newblock Time-dependent electron localization function.
\newblock {\em Phys. Rev. A}, 71(1):010501, 2005.

\end{thebibliography}
 
\end{document}